\documentclass[12 pt]{article}
\usepackage{amssymb}
\usepackage{amsfonts}
\usepackage{amsmath}
\usepackage{graphicx}
\usepackage{subcaption}
\usepackage{epstopdf}
\usepackage[font={small,it}]{caption}
\usepackage{hyperref}
\usepackage{cite}

\graphicspath{ {figure/} }
\catcode`\@=11

\@addtoreset{equation}{section} \catcode`\@=10

\begin{document}

\author{A. S.-F. Obada$^{1}$, M. M. A. Ahmed$^{1}$\ , M. Abu-Shady$^{2}$and
H. F. Habeba$^{2}$ \\
%EndAName
Department of Mathematics, Faculty of Science, Al-Azher University,\\
Nassr City 11884, Egypt$^{1}$\\
Department of Mathematics and Computer Science, Faculty of Science, \\
Menoufia University, Shebin Elkom 32511, Egypt$^{^{2}}$ \ \ \ \ \ \ }
\title{ Effect of an External Magnetic Field on Some Statistical Properties
of the 2+1 Dirac-Moshinsky Oscillator }
\date{}
\maketitle

\begin{abstract}
The 2+1 Dirac-Moshinsky oscillator ( 2+1 DMO ) is mapped into the
generalized Jaynes-Cummings model ( GJCM ), in which an external magnetic
field is coupled to an external isospin field. The basic equations of model
are analytically solved, where the coherent state is considered as an
initial state. The obtained results show that the strength of the magnetic
field and the coupling parameter of the isospin field play important roles
on some statistical properties such as entanglement, population inversion
and degree of coherence. It has been shown that these parameters play a rule
to increase entanglement and show the collapses and revivals phenomenon.
\end{abstract}

 \newpage

\section{Introduction}

In quantum optics, the JCM is composed of a single two-level particle
interacting with a single quantized cavity mode of the electromagnetic field 
\cite{1}. This model is exactly solvable in the rotating wave approximation
and experimentally realized \cite{2}. It has been found that the JCM has
some statistical properties that are not in classical fields, such as the
degree of coherence, the collapses and revivals phenomenon and squeezing 
\cite{3,4}.

JCM has been used to elucidate the strong quantum correlation
(entanglement), which is an important aspect of quantum systems. It
demonstrates correlations, that cannot be discussed classically \cite{5}.

The Dirac oscillator was suggested \cite{6,7} and reinvestigated where the
linear term $im\omega c\beta \alpha .r$ is added to the relativistic
momentum of the free-particle Dirac equation \cite{8,9}.

The 1+1 Dirac-Moshinsky oscillator (1+1 DMO) has been exactly solved by
using the theory of the non-relativistic harmonic oscillator \cite{10,11}.
The Dirac oscillator has attracted a lot of attention and found many
applications in different branches of physics \cite{12,13,14,15}. In the
other hand the 2+1 dimensions have been related to quantum optics via the
JCM \cite{15,16}.

One of the most exciting properties of the DMO is its connection to quantum
optics \cite{17,18}. The connection to the quantum optics allows to conceive
quantum optics experiments that emulate this system. The dynamics of the 2+1
DMO were studied \cite{19}, in which an exact mapping of this quantum
relativistic system into the JCM is obtained. In \cite{20}, the 2+1 DMO in
the presence of an external magnetic field has been studied. Also, the
connection between anti-JCM with the DMO in a magnetic field was established
without making any limit on the strength of the magnetic field.

The 1+1 and 2+1 DMO have been mapped to the JCM and the dynamical features
of a Dirac particle under the influence of the external field have been
studied only at vacuum state \cite{21}. The previous attempts \cite{21}
concenterated on the number state without using the coherent state, where
the external isospin field is included only, also in Ref. \cite{19} the 2+1
DMO in an external magnetic field has been studied without any study of
statistical properties of the system.

In \cite{22} the 2+1 DMO coupled to an external field has been mapped into
the JCM . The effect of both the detuning parameter and the coherence angle
on the entanglement and the population inversion have been studied by using
two cases for the initiall state: the number state and the coherent state.
It has been shown that the coherent state gives good description for the
entanglement and the population inversion.

In this paper, we study the Dirac oscillator coupled to an isospin field in
the presence of an external magnetic field by mapping it to the GJCM. The
wave function is obtained by using the coherent state as an initial state.
In addition, we study the influence of both the strength of the magnetic
field and the coupling parameter of the isospin field on some non-classical
properties of the system.

This paper is arranged as follows: Sec. 2 is devoted to introduce basic
equations and relations. Sec. 3 is devoted to explain the mapping of 2+1 DMO
in an external magnetic field coupled to an external isospin field into the
GJCM. Sec. 4 is devoted to the analytical solution of the model, this is
followed by a discussion of some non-classical properties in Sec. 5.
Finally, in Sec. 6, we conclude this paper with some brief remarks.

\section{ Basic equations and relations}

\subsection{The 2+1 Dirac-Moshinsky Oscillator}

The DMO is introduced by Moshinsky and Szczepaniak \cite{8} by adding the
linear term $im\omega c\beta \alpha .r$ \ to the Dirac Hamiltonian for a
free particle. In the non-relativistic limit, it corresponds to the harmonic
oscillator plus a spin-orbit coupling term. The DMO model in 2+1 dimensions
takes the following form \cite{21} 
\begin{equation}
i\hbar \frac{\partial }{\partial t}\left\vert \psi \right\rangle
=[\sum\limits_{j=1}^{2}c\alpha _{_{j}}(p_{_{j}}+im\omega \beta
r_{j})+mc^{2}\beta ]\left\vert \psi \right\rangle ,  \label{1}
\end{equation}

where $c$ is the speed of light, $m$ is the rest mass of the particle. $%
\alpha _{_{j}}$ and $\beta $ are the Dirac matrices in the standard
representation, they are taken here as $\alpha _{_{1}}=-\hat{\sigma}_{y},$ $%
\alpha _{_{2}}=-\hat{\sigma}_{x}\ $and $\beta =\hat{\sigma}_{z},$ \ where
the $\hat{\sigma}$'s are the Pauli matrices. $\omega $ represents the
harmonic oscillator frequency. We note that the standard Dirac equation is
recovered at $\omega =0$ \cite{23}.

The 2+1 DMO in an external magnetic field takes the following form \cite{19}

\begin{equation}
\hat{H}^{2}=\sum\limits_{j=1}^{2}c\alpha _{_{j}}(p_{_{j}}-\frac{e\mathbf{A}%
_{j}}{c}-im\omega \beta r_{j})+mc^{2}\beta ,  \label{121}
\end{equation}

where e is the charge of the DMO and \textbf{A} is the vector potential. We
take the magnetic field \textbf{B }in the z-direction, so the vector
potential for this particular magnetic field takes the form $\mathbf{A=(}%
\frac{-B}{2}y,\frac{B}{2}x,0\mathbf{),}$ or $\mathbf{A=}\frac{1}{2}(\mathbf{%
B\wedge r}).$

\subsection{The generalized Jaynes-Cummings model}

We briefly introduce the GJCM, in order to study and connect it with more
general and complicated systems, besides the Dirac oscillator.

\ It is a theoretical model in quantum optics. It describes the system of a
two-level particle interacting with one mode of the electromagnetic field
without using the rotating wave approximation. The Hamiltonian in the
interaction picture takes the following form \cite{24} 
\begin{equation}
\hat{H}_{JC}=\Omega (\hat{\sigma}_{+}+\hat{\sigma}_{-})(\hat{a}+\hat{a}%
^{\dagger })+\delta \hat{\sigma}_{z},  \label{2}
\end{equation}%
where $\Omega $ is the particle-field coupling constant, the operators $\hat{%
\sigma}_{+}$ and $\hat{\sigma}_{-}$ are the raising and lowering operators
for the two-level system, they satisfy the commutation relations $\left[ 
\hat{\sigma}_{z},\hat{\sigma}_{\pm }\right] =\pm 2\hat{\sigma}_{\pm }$ and $%
\left[ \hat{\sigma}_{+},\hat{\sigma}_{-}\right] =\hat{\sigma}_{z}$. $\hat{a}%
^{\dagger }$ and $\hat{a}$ are the Boson creation and annihilation operator
respectively which satisfy the commutation relation $[\hat{a},\hat{a}%
^{\dagger }]=1.$ $\delta $ stands for the detuning of the atomic transition
frequency from the cavity mode frequency. \bigskip

\section{ Mapping of the 2+1 DMO model in an external magnetic field coupled
to an external isospin field into the GJCM}

By using the spinor $\left\vert \psi \right\rangle =%
\begin{bmatrix}
\left\vert \psi _{1}\right\rangle \\ 
\left\vert \psi _{2}\right\rangle%
\end{bmatrix}%
$ and $\hat{H}^{2}\left\vert \psi \right\rangle =E\left\vert \psi
\right\rangle ,$ Eq. (\ref{121}) becomes a set of coupled equations as
follows%
\begin{equation}
(E-mc^{2})\left\vert \psi _{1}\right\rangle =(2cp_{z}+im\tilde{\omega}\bar{z}%
)\left\vert \psi _{2}\right\rangle  \label{3}
\end{equation}%
\begin{equation}
(E+mc^{2})\left\vert \psi _{2}\right\rangle =(2cp_{\bar{z}}-im\tilde{\omega}%
z)\left\vert \psi _{1}\right\rangle ,  \label{4}
\end{equation}

where 
\begin{eqnarray}
p_{z} &=&\frac{1}{2}[p_{x}-ip_{y}],p_{\bar{z}}=\frac{1}{2}[p_{x}+ip_{y}],
\label{5} \\
z &=&(x+iy),\bar{z}=(x-iy),  \label{6} \\
\tilde{\omega} &=&\omega +\frac{\omega _{c}}{2},  \label{7}
\end{eqnarray}

with $\omega _{c}=\frac{-\left\vert eB\right\vert }{mc}$ is the cyclotron
frequency.

We can write $\hat{H}^{2}$ in the following matrix form

\begin{equation}
\hat{H}^{2}=%
\begin{pmatrix}
mc^{2} & 2cp_{z}+imc\tilde{\omega}\bar{z} \\ 
2cp_{\bar{z}}-imc\tilde{\omega}z & -mc^{2}%
\end{pmatrix}%
.  \label{8}
\end{equation}

We note that the 2+1 DMO with angular frequency $\omega $ in the presence of
the magnetic field maps into 2+1 DMO where the angular frequency $\omega $
changes to $\tilde{\omega}=\omega +\frac{\omega _{c}}{2},$ which means the
magnetic field decreases the angular frequency by half of the cyclotron
frequency of this system.

In order to find the solution, we define the following creation and
annihilation operators: 
\begin{eqnarray}
\hat{a} &=&\frac{1}{\sqrt{m\tilde{\omega}\hbar }}p_{\bar{z}}-\frac{i}{2}%
\sqrt{\frac{m\tilde{\omega}}{\hbar }}z,  \label{9} \\
\hat{a}^{\dagger } &=&\frac{1}{\sqrt{m\tilde{\omega}\hbar }}p_{z}+\frac{i}{2}%
\sqrt{\frac{m\tilde{\omega}}{\hbar }}\bar{z},  \label{10}
\end{eqnarray}

where $\left[ \hat{a},\hat{a}^{\dagger }\right] =1,\left[ \hat{a},\hat{a}%
\right] =0=\left[ \hat{a}^{\dagger },\hat{a}^{\dagger }\right] .$

Now, we can write the Hamiltonian $\hat{H}^{2}$ (\ref{8}) in terms of the
creation and annihilation operators as follows: 
\begin{equation}
\hat{H}^{2}=\eta (\hat{a}^{\dagger }\hat{\sigma}_{+}+\hat{a}\hat{\sigma}%
_{-})+mc^{2}\hat{\sigma}_{z},  \label{11}
\end{equation}

where $\eta =2\sqrt{mc^{2}\tilde{\omega}\hbar }.$ This equation represents
the Hamiltonian of the Anti-JCM in quantum optics.

In the presence of an external isospin field $\Phi $, the dynamics of the
total system is given by the Hamiltonian 
\begin{equation}
\tilde{H}=\hat{H}^{2}+\Phi ,  \label{12}
\end{equation}%
where $\hat{H}^{2}$ is given by Eq. (\ref{11}) and $\Phi $ is the hermitean
operator. It takes the following form \cite{25}

\begin{equation}
\Phi =(A+\hat{\sigma}_{z}B)(\hat{a}\acute{\sigma}_{+}+\hat{a}^{\dagger }%
\acute{\sigma}_{-}+\gamma \acute{\sigma}_{z}),  \label{13}
\end{equation}%
where $\acute{\sigma}^{,}$s are the vectors of Pauli matrices, they have the
same commutation relations as $\hat{\sigma}^{,}$s, so, the corresponding
ladder operators are defined by 
\begin{equation}
\acute{\sigma}_{\pm }=\frac{1}{2}(\acute{\sigma}_{x}\pm i\acute{\sigma}_{y}).
\label{14}
\end{equation}

We use the simplest\ form of $\Phi $ (i.e.linear) as 
\begin{equation}
\Phi =\chi (\hat{a}\grave{\sigma}_{+}+\hat{a}^{\dagger }\grave{\sigma}%
_{-})+\gamma \grave{\sigma}_{z}.  \label{15}
\end{equation}

\bigskip $\tilde{H}$ can be described in quantum optics as GJCM, where $\hat{%
a}$ is the annihilation operator of the cavity field and each isospin with
an atom, while $\eta $ and $\chi $ can be described as the coupling of each
atom to the cavity isospin and $mc^{2}$ and $\gamma $ are described as the
detuning of each transition level with the cavity mode frequency. This model
can be seen as a linear combinition of the two JCM. This model may be
considered general than the model in \cite{22}, where in our model we take
into account the influence of the magnetic field, which lead to map this
model into the GJCM.

In order to solve this system (\ref{12}), we use the Heisenberg equation of
motian to deduce the constant of motion as follow 
\begin{equation}
I=\hat{n}+\frac{1}{2}(\grave{\sigma}_{z}-\hat{\sigma}_{z}),  \label{16}
\end{equation}

with $\hat{n}=\hat{a}^{\dagger }\hat{a}.$

\section{The analytical solution}

This section is devoted to derive the wave function $\left\vert \psi
(t)\right\rangle $ and the reduced density operators. We assume the two
particles (the particle in DMO and the isospin field) and the
electromagnetic field are initially prepared in ground states and coherent
state respectively. In this case, the wave function of this system at $t=0$
can be written as 
\begin{equation}
\left\vert \psi (0)\right\rangle =\left\vert -\right\rangle _{Ds}\otimes
\left\vert \grave{-}\right\rangle _{Is}\otimes \left\vert \alpha
\right\rangle _{F},  \label{17}
\end{equation}

where%
\begin{equation}
\left\vert \alpha \right\rangle =\sum_{n=0}^{\infty }q_{n}\left\vert
n\right\rangle ,  \label{18}
\end{equation}

with 
\begin{equation}
q_{n}=\exp (\frac{-\left\vert \alpha \right\vert ^{2}}{2})\frac{\alpha ^{n}}{%
\sqrt{n!}},\alpha \in 
%TCIMACRO{\U{2102} }%
%BeginExpansion
\mathbb{C}
%EndExpansion
.  \label{19}
\end{equation}

By using the constant of motion Eq. (\ref{16}), the wave function $%
\left\vert \psi (t)\right\rangle $ takes the following form at $t>0$%
\begin{eqnarray}
\left\vert \psi (t)\right\rangle &=&\sum\limits_{n=0}^{\infty
}(B_{1}(n,t)\left\vert -\grave{-},n+2\right\rangle +B_{2}(n,t)\left\vert +%
\grave{-},n+3\right\rangle  \notag \\
&&+B_{3}(n,t)\left\vert -\grave{+},n+1\right\rangle +B_{4}(n,t)\left\vert +%
\grave{+},n+2\right\rangle ).  \label{20}
\end{eqnarray}%
We obtain the coefficients $B_{j}(n,t),$ $(j=1,2,3,4)$ by solving the Schr%
\"{o}dinger equation. Therefore, we have the following system of
differential equations for the $B_{j}(n,t)$ coefficients 
\begin{equation}
i\hbar \dot{B}_{1}(n,t)=a(n)B_{2}(n,t)+d(n)B_{3}(n,t)-2\Omega B_{1}(n,t),
\label{21}
\end{equation}%
\begin{equation}
i\hbar \dot{B}_{2}(n,t)=a(n)B_{1}(n,t)+c(n)B_{4}(n,t),  \label{22}
\end{equation}%
\begin{equation}
i\hbar \dot{B}_{3}(n,t)=b(n)B_{4}(n,t)+d(n)B_{1}(n,t),  \label{23}
\end{equation}%
\begin{equation}
i\hbar \dot{B}_{4}(n,t)=b(n)B_{3}(n,t)+c(n)B_{2}(n,t)+2\Omega B_{1}(n,t),
\label{24}
\end{equation}

where%
\begin{equation}
a(n)=\lambda _{1}\sqrt{n+3},b(n)=a(n-1),  \label{25}
\end{equation}%
\begin{equation}
c(n)=\lambda _{2}\sqrt{n+3},d(n)=c(n-1),  \label{26}
\end{equation}%
\begin{equation}
mc^{2}=\gamma =\Omega ,  \label{27}
\end{equation}

with%
\begin{equation*}
\lambda _{1}=\frac{2\sqrt{1+\xi }}{\eta },\lambda _{2}=\frac{\chi }{\eta 
\sqrt{mc^{2}\omega }}\ and\text{ }\xi =\frac{eB}{2mc\omega }.
\end{equation*}

We take $c=1=\hbar $. The time-dependent coefficients $B_{j}(n,t),$ $%
(j=1,2,3,4)$ are obtained, by solving the above differential equations (\ref%
{21}-\ref{24}). With the wave function $\left\vert \psi (t)\right\rangle $
calculated, then calculations for any property related to the particles or
the field can be performed.

The reduced density operator of the isospin field $\grave{\rho}(t)$ can be
obtained as following

\begin{eqnarray}
\grave{\rho}(t) &=&Tr_{F}Tr_{DO}\left\vert \psi (t)\right\rangle
\left\langle \psi (t)\right\vert   \notag \\
&=&\grave{\rho}_{ee}(t)\left\vert \grave{+}\right\rangle \left\langle \grave{%
+}\right\vert +\grave{\rho}_{gg}(t)\left\vert \grave{-}\right\rangle
\left\langle \grave{-}\right\vert +  \notag \\
&&\grave{\rho}_{eg}(t)\left\vert \grave{+}\right\rangle \left\langle \grave{-%
}\right\vert +\grave{\rho}_{ge}(t)\left\vert \grave{-}\right\rangle
\left\langle \grave{+}\right\vert ,  \label{28}
\end{eqnarray}

where%
\begin{eqnarray}
\grave{\rho}_{ee}(t) &=&\sum\limits_{n=0}^{\infty }(\left\vert
B_{3}(n,t)\right\vert ^{2}+\left\vert B_{4}(n,t)\right\vert ^{2}),
\label{29} \\
\grave{\rho}_{gg}(t) &=&\sum\limits_{n=0}^{\infty }(\left\vert
B_{1}(n,t)\right\vert ^{2}+\left\vert B_{2}(n,t)\right\vert ^{2}),
\label{30} \\
\grave{\rho}_{eg}(t) &=&\sum\limits_{n=0}^{\infty }(B_{3}(n+1,t)B_{1}^{\ast
}(n,t)+B_{4}(n+1,t)B_{2}^{\ast }(n,t))=\grave{\rho}_{ge}^{\ast }(t).
\label{31}
\end{eqnarray}

Also, to obtain the reduced density matrix of the two particles, we trace
over the oscillator degree of freedom 
\begin{eqnarray}
\rho (t) &=&Tr_{F}\left\vert \psi (t)\right\rangle \left\langle \psi
(t)\right\vert   \notag \\
&=&%
\begin{pmatrix}
\rho _{11}(t) & \rho _{12}(t) & \rho _{13}(t) & \rho _{14}(t) \\ 
\rho _{21}(t) & \rho _{22}(t) & \rho _{23}(t) & \rho _{24}(t) \\ 
\rho _{31}(t) & \rho _{32} & \rho _{33}(t) & \rho _{34}(t) \\ 
\rho _{41}(t) & \rho _{42} & \rho _{43}(t) & \rho _{44}(t)%
\end{pmatrix}%
,  \label{44}
\end{eqnarray}

where%
\begin{eqnarray}
\rho _{11}(t) &=&\sum\limits_{n=0}^{\infty }\left\vert B_{1}(n,t)\right\vert
^{2},\rho _{22}(t)=\sum\limits_{n=0}^{\infty }\left\vert
B_{2}(n,t)\right\vert ^{2},  \notag \\
\rho _{33}(t) &=&\sum\limits_{n=0}^{\infty }\left\vert B_{3}(n,t)\right\vert
^{2},\rho _{44}(t)=\sum\limits_{n=0}^{\infty }\left\vert
B_{4}(n,t)\right\vert ^{2},  \notag \\
\rho _{12}(t) &=&\sum\limits_{n=0}^{\infty }B_{1}(n+1,t)B_{2}^{\ast
}(n,t)=\rho _{21}^{\ast }(t),  \notag \\
\rho _{13}(t) &=&\sum\limits_{n=0}^{\infty }B_{1}(n,t)B_{3}^{\ast
}(n+1,t)=\rho _{31}^{\ast }(t),  \notag \\
\rho _{14}(t) &=&\sum\limits_{n=0}^{\infty }B_{1}(n,t)B_{4}^{\ast
}(n,t)=\rho _{41}^{\ast }(t),  \notag \\
\rho _{23}(t) &=&\sum\limits_{n=0}^{\infty }B_{2}(n,t)B_{3}^{\ast
}(n+2,t)=\rho _{32}^{\ast }(t),  \notag \\
\rho _{34}(t) &=&\sum\limits_{n=0}^{\infty }B_{3}(n+1,t)B_{4}^{\ast
}(n,t)=\rho _{43}^{\ast }(t).  \label{45}
\end{eqnarray}

With these operators given, different statistical properties of this system
can be studied.

\section{Non-classical properties}

In this section, we discuss some statistical properties of the present
system, where we concentrate on the influence of the strength of the
magnetic field and the coupling constant parameter of the isospin field on
the behaviour of the entanglement and the population inversion and the
correlation function.

\subsection{ Entanglement}

In this subsection we study the entanglement between the DMO and the isospin
field through the von Neumann entropy. In quantum optics the von Neumann
entropy has been used to study the dynamic charactertics of a two-level atom
interacting with light \cite{26}. It is noted that this measure is a useful
physical quantity for measuring the degree of entanglement in a pure state.

The von Neumann entropy is defined in quantum mechanics as \cite{22,27}, 
\begin{equation}
S(t)=-\lambda _{-}(t)\ln \lambda _{-}(t)-\lambda _{+}(t)\ln \lambda _{+}(t),
\label{38}
\end{equation}

where $\lambda _{\pm }(t)$ are the eigenvalues of the reduced density matrix 
$\grave{\rho}(t)$ Eqs. (\ref{28}- \ref{31}). They can be easily evaluated
through the following form:%
\begin{equation}
\lambda _{\pm }(t)=\frac{1}{2}\pm \frac{1}{2}\sqrt{\left\langle \hat{\sigma}%
_{x}(t)\right\rangle ^{2}+\left\langle \hat{\sigma}_{y}(t)\right\rangle
^{2}+\left\langle \hat{\sigma}_{z}(t)\right\rangle ^{2}},  \label{39}
\end{equation}%
where

\begin{eqnarray}
\left\langle \hat{\sigma}_{x}(t)\right\rangle &=&2\Re \lbrack \grave{\rho}%
_{eg}(t)],  \label{40} \\
\left\langle \hat{\sigma}_{y}(t)\right\rangle &=&2\Im \lbrack \grave{\rho}%
_{eg}(t)],  \label{41} \\
\left\langle \hat{\sigma}_{z}(t)\right\rangle &=&\grave{\rho}_{ee}(t)-\grave{%
\rho}_{gg}(t).  \label{42}
\end{eqnarray}

\begin{figure}[h]
\begin{subfigure}{1\textwidth}
		\begin{subfigure}{0.50\textwidth}
			\includegraphics[width=\linewidth, height=5cm]{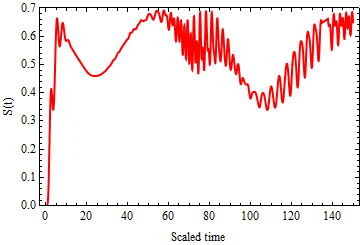}
			\caption{$\lambda_{1}=0.2\lambda$}
		\end{subfigure}
		\begin{subfigure}{0.50\textwidth}
			\includegraphics[width=\linewidth, height=5cm]{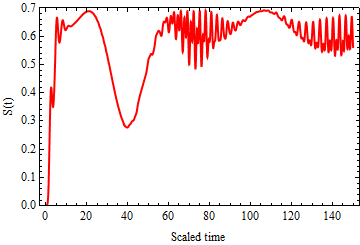}
			\caption{$\lambda_{1}=0.5\lambda$}
		\end{subfigure}
		
	\end{subfigure}
\newline
\begin{subfigure}{1\textwidth}
		\begin{subfigure}{0.50\textwidth}
			
			\includegraphics[width=\linewidth, height=5cm]{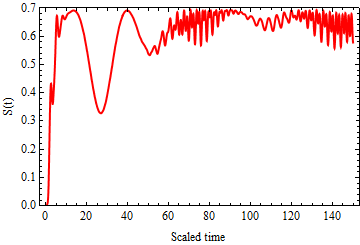}
			\caption{$\lambda_{1}=0.8\lambda$}
		\end{subfigure}
		\begin{subfigure}{0.50\textwidth}
			\includegraphics[width=\linewidth, height=5cm]{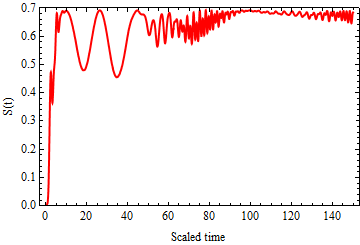}
			\caption{$\lambda_{1}=1.2\lambda$}
		\end{subfigure}

	\end{subfigure}
\caption{The von Neumann entropy is plotted as a function of $\protect%
\lambda t$ with $\Omega =0.2\protect\lambda $ and $\protect\lambda _{2}=0.3%
\protect\lambda .$}
\label{fig:3}
\end{figure}

\begin{figure}[h]
\begin{subfigure}{1\textwidth}
		\begin{subfigure}{0.50\textwidth}
			\includegraphics[width=\linewidth, height=5cm]{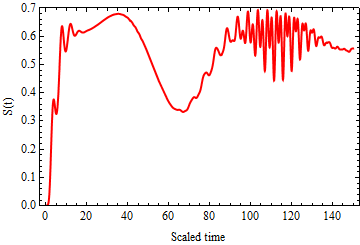}
			\caption{$\lambda_{2}=0.2\lambda$}
		\end{subfigure}
		\begin{subfigure}{0.50\textwidth}
			\includegraphics[width=\linewidth, height=5cm]{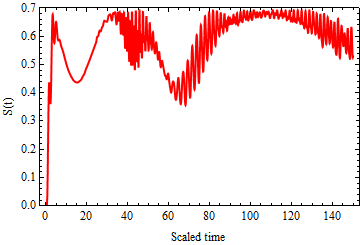}
			\caption{$\lambda_{2}=0.5\lambda$}
		\end{subfigure}
		
	\end{subfigure}
\newline
\begin{subfigure}{1\textwidth}
		\begin{subfigure}{0.50\textwidth}
			
			\includegraphics[width=\linewidth, height=5cm]{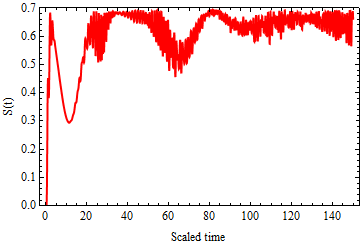}
			\caption{$\lambda_{2}=0.8\lambda$}
		\end{subfigure}
		\begin{subfigure}{0.50\textwidth}
			\includegraphics[width=\linewidth, height=5cm]{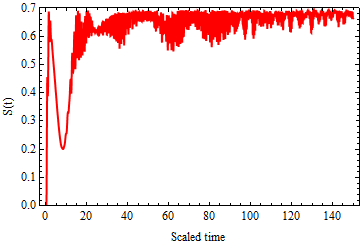}
			\caption{$\lambda_{2}=1.2\lambda$}
		\end{subfigure}

	\end{subfigure}
\caption{The von Neumann entropy is plotted as a function of $\protect%
\lambda t$ with $\Omega =0.2\protect\lambda $ and $\protect\lambda _{1}=0.3%
\protect\lambda .$}
\label{fig:77}
\end{figure}

In Figs. (\ref{fig:3},\ref{fig:77}), we display the effect of the strength
of the magnetic field and the coupling parameter of the isospin field on the
evolution of the von Neumann entropy against the scaled time $\lambda t,$
where $\lambda =\eta \sqrt{mc^{2}\omega },$ when the atoms in ground state
initially and the field be prepared initially in the coherent state. The
value of the intensity of the initial coherent parameter has been fixed as $%
\alpha =3$ and the detuning parameter has been fixed as $\Omega =0.2\lambda
. $

We note that $S(t)$ starts from zero, then it followed by a sequence of
fluctuations in the oscillation. This means that this system begins by
disentangled state (at $\lambda t=0$) then it develops to a mixed state (at $%
\lambda t>0$) and never reachs to the pure state again.

In Fig. (\ref{fig:3}), the effect of the strength of the magnetic field ($%
\lambda _{1}$) appear clearly where there is a sudden decrease in the value
of $S(t)$ as $\lambda t$ ranges from $20$ to $40,$ see Fig. (\ref{fig:3}a).
Also an extra minimum (decrease of $S(t)$) occurs as $\lambda _{1}$
increases at the same period Fig. (\ref{fig:3}c,\ref{fig:3}d), also the
entanglement increases for a longer period ($\lambda t>40$). By increasing $%
\lambda _{1}$, $S(t)$ oscillates near the maximum value of ($\ln 2$).

Fig. (\ref{fig:77}) shows the effect of the coupling parameter of the
isospin field ($\lambda _{2}$) on the entanglement, we observe that by
increasing the value of this parameter, the value of entanglement increases
also the number of the fluctuation increases.

We may conclude that to obtain strong entanglement between the isospin field
and the Dirac oscillator, we increase the value of $\lambda _{1}$ or $%
\lambda _{2}$.


\subsection{Concurrence}

In this subsection, we use the concurrence to measure the entanglement
between the two particles. It ensures the scale between $0$ for a separable
(disentangled) state and $\sqrt{2(N-1)\diagup N}$ for the maximally
entangled state. The concurrence may be written in the following form \cite%
{28,29}:

\begin{equation*}
C(t)=\sqrt{2\sum_{i,j=1,2,3,4}(\rho _{ii}(t)\rho _{jj}(t)-\rho _{ij}(t)\rho
_{ji}(t))},i\neq j,
\end{equation*}

where $\rho _{ii}(t),\rho _{jj}(t),\rho _{ji}(t)$ and $\rho _{ij}(t)$ are
given by (\ref{45}).

\begin{figure}[h]
\begin{subfigure}{1\textwidth}
		\begin{subfigure}{0.50\textwidth}
			\includegraphics[width=\linewidth, height=5cm]{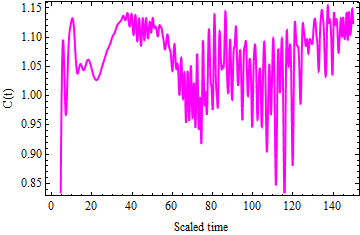}
			\caption{$\lambda_{1}=0.2\lambda$}
		\end{subfigure}
		\begin{subfigure}{0.50\textwidth}
			\includegraphics[width=\linewidth, height=5cm]{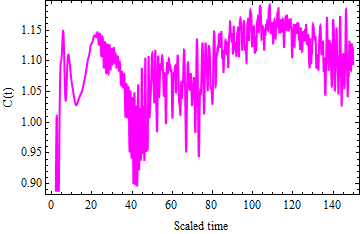}
			\caption{$\lambda_{1}=0.5\lambda$}
		\end{subfigure}
		
	\end{subfigure}
\newline
\begin{subfigure}{1\textwidth}
		\begin{subfigure}{0.50\textwidth}
			
			\includegraphics[width=\linewidth, height=5cm]{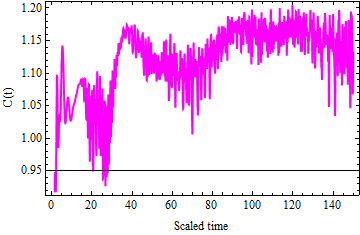}
			\caption{$\lambda_{1}=0.8\lambda$}
		\end{subfigure}
		\begin{subfigure}{0.50\textwidth}
			\includegraphics[width=\linewidth, height=5cm]{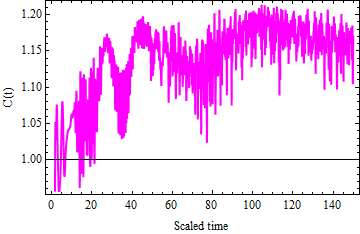}
			\caption{$\lambda_{1}=1.2\lambda$}
		\end{subfigure}

	\end{subfigure}
\caption{The concurrence is plotted as a function of $\protect\lambda t$ .
The parameters are similar to \protect\ref{fig:3}.}
\label{fig:7}
\end{figure}

\begin{figure}[h]
\begin{subfigure}{1\textwidth}
		\begin{subfigure}{0.50\textwidth}
			\includegraphics[width=\linewidth, height=5cm]{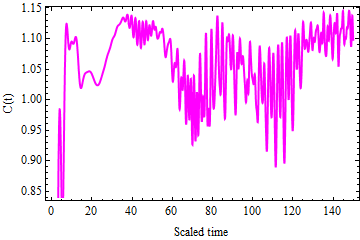}
			\caption{$\lambda_{2}=0.2\lambda$}
		\end{subfigure}
		\begin{subfigure}{0.50\textwidth}
			\includegraphics[width=\linewidth, height=5cm]{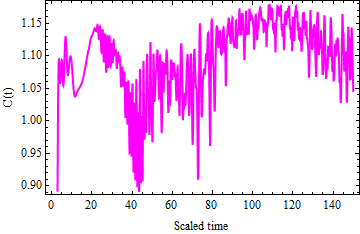}
			\caption{$\lambda_{2}=0.5\lambda$}
		\end{subfigure}
		
	\end{subfigure}
\newline
\begin{subfigure}{1\textwidth}
		\begin{subfigure}{0.50\textwidth}
			
			\includegraphics[width=\linewidth, height=5cm]{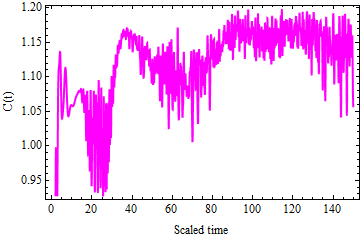}
			\caption{$\lambda_{2}=0.8\lambda$}
		\end{subfigure}
		\begin{subfigure}{0.50\textwidth}
			\includegraphics[width=\linewidth, height=5cm]{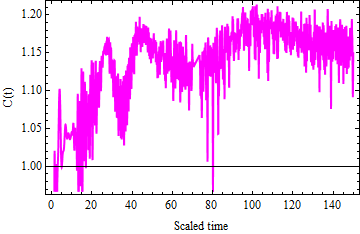}
			\caption{$\lambda_{2}=1.2\lambda$}
		\end{subfigure}

	\end{subfigure}
\caption{The concurrence is plotted as a function of $\protect\lambda t$ .
The parameters are similar to \protect\ref{fig:77}.}
\label{fig:11}
\end{figure}

In Figs. (\ref{fig:7}, \ref{fig:11}), we plot the evolution of concurrence $%
C(t)$ versus the scaled time $\lambda t$, in order to see the effect of $%
\lambda _{1}$ and $\lambda _{2}$ on the degree of the entanglement between
the two particles (the isospin field and the particle in DMO). We use the
same initial parameters as the previous figures.

It is noted that $C(t)$ starts from zero, then it is followed by a sequence
of fluctuations between $zero$ and $1.2,$ this means that the entanglement
between the two particles can not be performed before the interaction is
switched on. To visualize the effect of $\lambda _{1},$ see Fig. (\ref{fig:7}%
) where we take different values of $\lambda _{1}.$ It is observed that for
a large effect of $\lambda _{1},$ the entanglement increases after a short
time from the start, and the number of rapid fluctuations increases, see
Figs. (\ref{fig:7}c, \ref{fig:7}d).

The same behaviour appears in Fig. (\ref{fig:11}), where we use different
values of $\lambda _{2}.$ We can say that the effect of $\lambda _{1}$ on
the degree of the entanglement between the two particles is similar to the
effect of $\lambda _{2},$ where, by increasing the value of any of these
parameters, the degree of entanglement increases. 

\subsection{The population inversion}

The population inversion gives us information about the behaviour of the
particle during the interaction period, which determines when this particle
reaches its maximal state and leads one to observe when this particle is in
its excited or ground state or in a superposition state. From mathematical
point of view the population inversion is the expectation value of the
operator $\hat{\sigma}_{z}$, thus we have%
\begin{equation}
W(t)=\grave{\rho}_{ee}(t)-\grave{\rho}_{gg}(t).  \label{43}
\end{equation}

\begin{figure}[h]
\begin{subfigure}{1\textwidth}
		\begin{subfigure}{0.50\textwidth}
			\includegraphics[width=\linewidth, height=5cm]{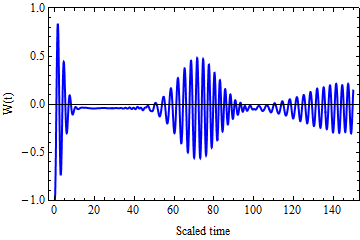}
			\caption{$\lambda_{1}=0.2\lambda$}
		\end{subfigure}
		\begin{subfigure}{0.50\textwidth}
			\includegraphics[width=\linewidth, height=5cm]{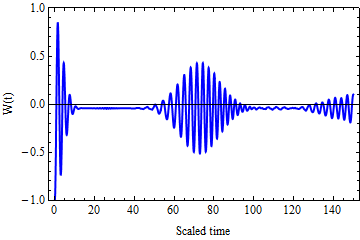}
			\caption{$\lambda_{1}=0.5\lambda$}
		\end{subfigure}
		
	\end{subfigure}
\newline
\begin{subfigure}{1\textwidth}
		\begin{subfigure}{0.50\textwidth}
			
			\includegraphics[width=\linewidth, height=5cm]{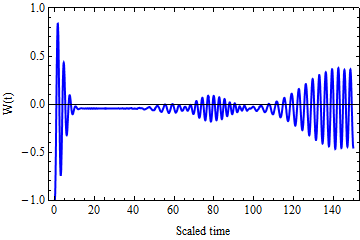}
			\caption{$\lambda_{1}=0.8\lambda$}
		\end{subfigure}
		\begin{subfigure}{0.50\textwidth}
			\includegraphics[width=\linewidth, height=5cm]{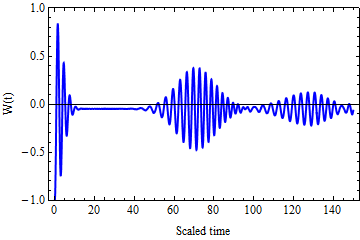}
			\caption{$\lambda_{1}=1.2\lambda$}
		\end{subfigure}

	\end{subfigure}
\caption{The population inversion of the isospin field is plotted as a
function of $\protect\lambda t$ . The parameters are similar to \protect\ref%
{fig:3}.}
\label{fig:2}
\end{figure}

\begin{figure}[h]
\begin{subfigure}{1\textwidth}
		\begin{subfigure}{0.50\textwidth}
			\includegraphics[width=\linewidth, height=5cm]{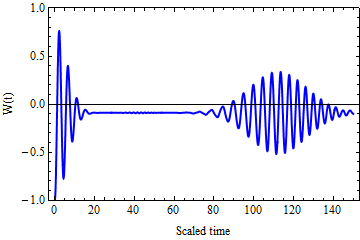}
			\caption{$\lambda_{2}=0.2\lambda$}
		\end{subfigure}
		\begin{subfigure}{0.50\textwidth}
			\includegraphics[width=\linewidth, height=5cm]{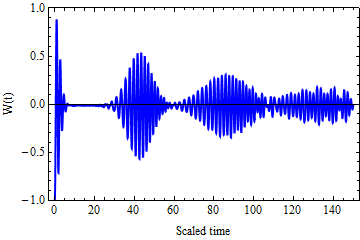}
			\caption{$\lambda_{2}=0.5\lambda$}
		\end{subfigure}
		
	\end{subfigure}
\newline
\begin{subfigure}{1\textwidth}
		\begin{subfigure}{0.50\textwidth}
			
			\includegraphics[width=\linewidth, height=5cm]{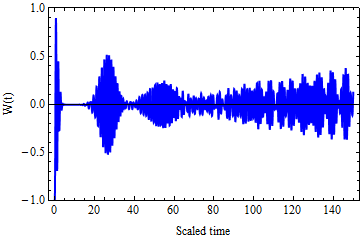}
			\caption{$\lambda_{2}=0.8\lambda$}
		\end{subfigure}
		\begin{subfigure}{0.50\textwidth}
			\includegraphics[width=\linewidth, height=5cm]{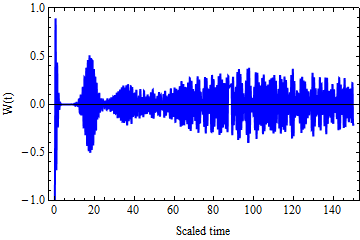}
			\caption{$\lambda_{2}=1.2\lambda$}
		\end{subfigure}

	\end{subfigure}
\caption{The population inversion of the isospin field is plotted as a
function of $\protect\lambda t$ . The parameters are similar to \protect\ref%
{fig:77}.}
\label{fig:10}
\end{figure}

We display the evolution of the population inversion of the isospin field
for different values of $\lambda _{1}$ in Fig. (\ref{fig:2}) and we use
different values of $\lambda _{2}$ in Fig. (\ref{fig:10}). We use the same
initial parameters as the previous figures. We note that the collapses and
revivals phenomenon is very obvious in all figures, the function $W(t)$ is
symmetric around $W(t)=0$ and the population inversion oscillates between
(-1) and (+1).

We observe that the strength of the magnetic field does not affect strongly
on the behaviour of the isospin field, see Fig. (\ref{fig:2}). We see the
opposite in Fig. (\ref{fig:10}), where by increasing the values of $\lambda
_{2}$, the collapse period decreases and the oscillation increases during
the revival period, see Fig. (\ref{fig:10}b). By taking large value of $%
\lambda _{2}$, the oscillation increases rapidly and the collapses and
revivals phenomenon do not appear as clearly as before due its interference
between the patterns, see Figs. (\ref{fig:10}c, \ref{fig:10}d).

We can say that in order to study the behaviour of the isospin field and
show the collapses and revivals phenomenon clearly in this system, we simply
increase the value of $\lambda _{2}.$

\subsection{The Second-order coherence}

No doubt the examination of the second-order correlation function leads to
better understanding for the non-classical behaviour of the system. For this
reason, we devote the present subsection to discuss the behaviour of the
correlation function for the present system. The correlation function is
usually used to discuss the sub-Poissonian and super-Poissonian behaviour of
the photon distribution. By using this function, we can distinguish between
classical and non-classical behaviour of the system. The normalized
second-order correlation function is defined by \cite{24}%
\begin{equation}
g^{(2)}(t)=\frac{\left\langle \hat{a}^{\dag 2}\hat{a}^{2}\right\rangle }{%
\left\langle \hat{a}^{\dag }\hat{a}\right\rangle ^{2}}.  \label{46}
\end{equation}

A light field has a sub-Poissonian distribution if $g^{(2)}(t)<1$, which is
a non-classical effect and means that the probability of detecting an
incident pair of photons is less than it would be for a coherent field
described by the Poissonian distribution. On the other hand, light has
super-Poissonian distribution if $g^{(2)}(t)>1$, which is a classical
effect, and a Poissonian distribution of photon (standard for the coherent
state) if $g^{(2)}(t)=1$. In the meantime, the system displays thermal
statistics when $g^{(2)}(t)=2$ and super-thermal for $g^{(2)}(t)>2$. In
order to discuss the distribution of this system, we calculate the
expectation value of the quantities $\left\langle \hat{a}^{\dag 2}\hat{a}%
^{2}\right\rangle $ and $\left\langle \hat{a}^{\dag }\hat{a}\right\rangle
^{2}$ 
\begin{eqnarray}
&&\left\langle \hat{a}^{\dagger 2}\hat{a}^{2}\right\rangle =\left\langle 
\hat{n}(\hat{n}-1)\right\rangle =  \notag \\
&&\sum\limits_{n=0}^{\infty }[(n+1)(n+2)\left\vert B_{1}(n,t)\right\vert
^{2}+(n+2)(n+3)\left\vert B_{2}(n,t)\right\vert ^{2}  \notag \\
&&+n(n+1)\left\vert B_{3}(n,t)\right\vert ^{2}+(n+1)(n+2)\left\vert
B_{4}(n,t)\right\vert ^{2}].  \notag \\
&&  \label{47}
\end{eqnarray}

\begin{eqnarray}
\left\langle \hat{a}^{\dag }\hat{a}\right\rangle ^{2} &=&\left\langle \hat{n}%
\right\rangle ^{2}=  \notag \\
&&(\sum\limits_{n=0}^{\infty }[(n+2)\left\vert B_{1}(n,t)\right\vert
^{2}+(n+3)\left\vert B_{2}(n,t)\right\vert ^{2}+  \notag \\
&&(n+1)\left\vert B_{3}(n,t)\right\vert ^{2}+(n+2)\left\vert
B_{4}(n,t)\right\vert ^{2}])^{2}  \label{48}
\end{eqnarray}

By using Eqs. (\ref{46})-(\ref{48}) we can easily get $g^{(2)}(t)$.

\begin{figure}[h]
\begin{subfigure}{1\textwidth}
		\begin{subfigure}{0.50\textwidth}
			\includegraphics[width=\linewidth, height=5cm]{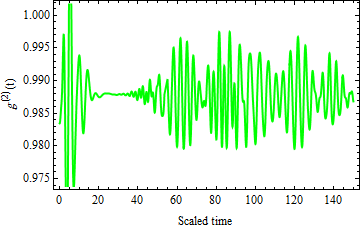}
			\caption{$\lambda_{1}=0.2\lambda$}
		\end{subfigure}
		\begin{subfigure}{0.50\textwidth}
			\includegraphics[width=\linewidth, height=5cm]{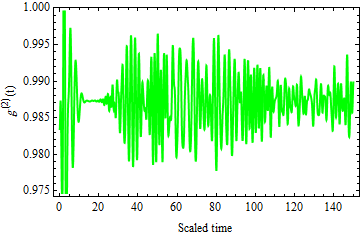}
			\caption{$\lambda_{1}=0.5\lambda$}
		\end{subfigure}
		
	\end{subfigure}
\newline
\begin{subfigure}{1\textwidth}
		\begin{subfigure}{0.50\textwidth}
			
			\includegraphics[width=\linewidth, height=5cm]{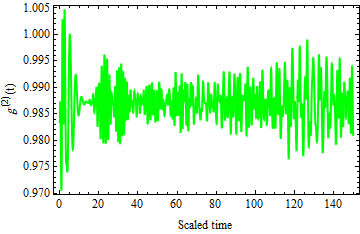}
			\caption{$\lambda_{1}=0.8\lambda$}
		\end{subfigure}
		\begin{subfigure}{0.50\textwidth}
			\includegraphics[width=\linewidth, height=5cm]{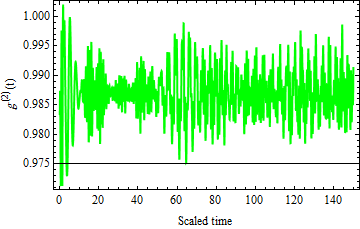}
			\caption{$\lambda_{1}=1.2\lambda$}
		\end{subfigure}

	\end{subfigure}
\caption{The second order correlation is plotted as a function of $\protect%
\lambda t$ with $.$ The parameters are similar to \protect\ref{fig:3}.}
\label{fig:8}
\end{figure}

\begin{figure}[h]
\begin{subfigure}{1\textwidth}
			\begin{subfigure}{0.50\textwidth}
				\includegraphics[width=\linewidth, height=5cm]{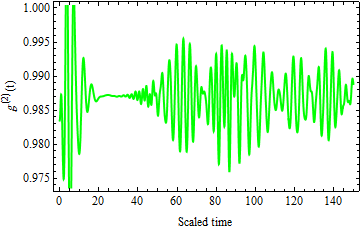}
				\caption{$\lambda_{2}=0.2\lambda$}
			\end{subfigure}
			\begin{subfigure}{0.50\textwidth}
				\includegraphics[width=\linewidth, height=5cm]{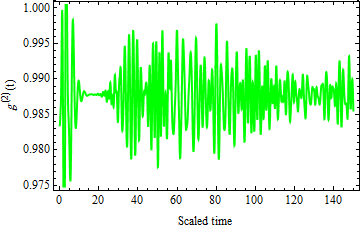}
				\caption{$\lambda_{2}=0.5\lambda$}
			\end{subfigure}
			
		\end{subfigure}
\newline
\begin{subfigure}{1\textwidth}
			\begin{subfigure}{0.50\textwidth}
				
				\includegraphics[width=\linewidth, height=5cm]{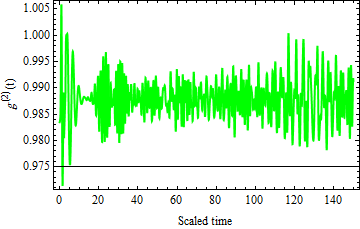}
				\caption{$\lambda_{2}=0.8\lambda$}
			\end{subfigure}
			\begin{subfigure}{0.50\textwidth}
				\includegraphics[width=\linewidth, height=5cm]{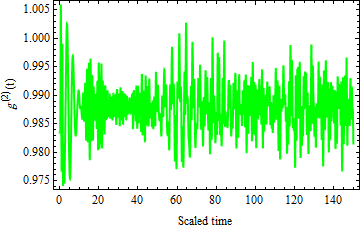}
				\caption{$\lambda_{2}=1.2\lambda$}
			\end{subfigure}

		\end{subfigure}
\caption{The second order correlation is plotted as a function of $\protect%
\lambda t$. The parameters are similar to \protect\ref{fig:77}.}
\label{fig:12}
\end{figure}

Now, we discuss the numerical calculations of the second-order correlation
function $g^{(2)}(t)$ in Figs. (\ref{fig:8}, \ref{fig:12}). It is observed
that the oscillation base line is oscilating around $0.987$ and never
reaches $1$ after $\lambda t>o,$  which means, the system is exhibiting
sub-Poissonian distribution, but the distribution is Poissonian at the
begining. Also, by increasing values of $\lambda _{1}$ or $\lambda _{2}$,\
the oscillation is squeezed and still sub-Poissonian. 

\section{\protect\bigskip Conclusion}

We have studied how the 2+1 DMO coupled to an external isospin field in an
external magnetic field is mapped into the GJCM. Also, we studied the effect
of the strength of the magnetic field ($\lambda _{1}$) and the coupling
parameter of the isospin field ($\lambda _{2}$) on the entanglement, the
population inversion and the second-order correlation function. We have used
the coherent state as the initial state and fixed the value of the initial
coherent parameter as $\alpha =3$. The model is considered more general than
the model obtained in \cite{22}, where we added to the 2+1 DMO an external
magnetic field which is neglected in \cite{22}. Also, this model is more
general than the model obtained in \cite{19}, where we use an external
isospin field and study some statistical properties of the system.

We would like to clarify that the strength of the magnetic field and the
coupling parameter have clear effects on the entanglement between the
isospin field and the Dirac oscillator, also, between the two particles.

The 2+1 DMO and the isospin field are separated at $\lambda t=0$ and they
are in a mixed state and never reach to the pure state again for any time $%
\lambda t>0$.

By increasing the value of $\lambda _{1}$ or the value of $\lambda _{2}$, we
can obtain strong entanglement between the isospin field and the Dirac
oscillator, see Figs. (\ref{fig:3}, \ref{fig:77}), also the degree of
entanglement between the two particles increases in a similar way, see Figs.
(\ref{fig:7}, \ref{fig:11}).

The behaviour of the isospin field and the collapses and revivals phenomenon
are shown clearly, by simply increasing the value of $\lambda _{2}$, see
Fig. (\ref{fig:10}).

The system is exhibiting sub-Poissonian distribution for any time $\lambda
t>0$.

So, this paper shows how important is the link between quantum optics and
quantum relativistic, where this link helps to study some of the statistical
properties of the 2+1 DMO coupled to an external isospin field in an
external magnetic field. These statistical properties have not been studied
without this link.

\end{document}